\documentclass[twocolumn,tighten]{aastex61}
\usepackage{natbib}
\usepackage{graphicx}
\usepackage{pslatex}
\newcommand{\Htwo}{\mbox{H$_{2}$}}
\newcommand{\Htwoo}{\mbox{H$_{2}$O}}
\newcommand{\Htwoco}{\mbox{H$_{2}$CO}}

\newcommand{\asig}{$A\,^2\Sigma ^{+}$}
\newcommand{\xsig}{$X\,^1\Sigma ^{+}$}
\newcommand{\api}{$A\,^1\Pi$}
\newcommand{\xpi}{$X\,^2\Pi$}

\newcommand{\kms}{km~s$^{-1}$}
\newcommand{\mols}{molecules~s$^{-1}$}
\newcommand{\iue}{\textit{IUE\/}}
\newcommand{\hst}{\textit{HST\/}}
\newcommand{\fuse}{\emph{FUSE\/}}

\shorttitle{Far-ultraviolet Spectroscopy of Comets with HST/COS}
\shortauthors{Feldman, et al.}

\begin{document}

\title{Far-ultraviolet Spectroscopy of Recent Comets with the Cosmic Origins Spectrograph on the Hubble Space Telescope} 

\correspondingauthor{Paul D. Feldman}
\email{pfeldman@jhu.edu}

\author[0000-0002-9318-259X]{Paul D. Feldman}
\affil{Department of Physics and Astronomy, The Johns Hopkins University, 3400 N. Charles Street, Baltimore, Maryland 21218}
 
\author{Harold A. Weaver}
\affil{Space Exploration Sector, Johns Hopkins University Applied Physics Laboratory, 11100 Johns Hopkins Road, Laurel, MD 20723-6099}

\author{Michael F. A'Hearn}
\altaffiliation{deceased}
\affil{Astronomy Department, University of Maryland, College Park, MD 20742}

\author{Michael R. Combi}
\affil{Department of Climate and Space Sciences and Engineering, University of Michigan, Ann Arbor, MI 48109}

\author{Neil Dello Russo}
\affil{Space Exploration Sector, Johns Hopkins University Applied Physics Laboratory, 11100 Johns Hopkins Road, Laurel, MD 20723-6099}

\pagestyle{myheadings}
\markright{\today}

\begin{abstract}

Since its launch in 1990, the {\it Hubble Space Telescope} (\hst) has served as a platform with unique capabilities for remote observations of comets in the far-ultraviolet region of the spectrum.  Successive generations of imagers and spectrographs have seen large advances in sensitivity and spectral resolution enabling observations of the diverse properties of a representative number of comets during the past 25 years.  To date, four comets have been observed in the far-ultraviolet by the Cosmic Origins Spectrograph (COS), the last spectrograph to be installed in \hst, in 2009:  103P/Hartley 2, C/2009 P1 (Garradd), C/2012 S1 (ISON), and C/2014 Q2 (Lovejoy).  COS has unprecedented sensitivity, but limited spatial information in its $2.\!''5$ diameter circular aperture, and our objective was to determine the CO production rates from measurements of the CO Fourth Positive system in the spectral range of 1400 to 1700 \AA.  In the two brightest comets, nineteen bands of this system were clearly identified.  The water production rates were derived from nearly concurrent observations of the OH (0,0) band at 3085 \AA\ by the Space Telescope Imaging Spectrograph (STIS).  The derived CO/\Htwoo\ production rate ratio ranged from $\sim$0.3\% for Hartley 2 to $\sim$22\% for Garradd.  In addition, strong partially resolved emission features due to multiplets of \ion{S}{1}, centered at 1429 \AA\ and 1479 \AA, and of \ion{C}{1} at 1561 \AA\ and 1657 \AA, were observed in all four comets.  Weak emission from several lines of the \Htwo\ Lyman band system, excited by solar Ly$\alpha$ and Ly$\beta$ pumped fluorescence, were detected in comet Lovejoy.

\end{abstract}


\section{INTRODUCTION}

We have previously commented on the extremely high sensitivity of the Cosmic Origins Spectrograph (COS), installed in \hst\ in May 2009, to detect the far-ultraviolet CO Fourth Positive system (\api\ -- \xsig) in comets with CO abundance relative to water of less than 1\%.  That measurement, of comet 103P/Hartley 2 at the time of the EPOXI fly-by in November 2010, gave a relative abundance of $\sim$0.3\% \citep{Weaver:2011}.  In January 2012 we activated an \hst\ target-of-opportunity program to observe the active comet C/2009 P1 (Garradd) with COS to measure the CO abundance, in addition to observing with the Space Telescope Imaging Spectrograph (STIS) to pursue a number of other objectives.  This comet was discovered at a heliocentric distance of 8.7~au, reached perihelion at 1.55~au on 2011 Dec. 23.67, which enabled measurements of its evolution over an extended range of time and heliocentric distance.  Extensive observing campaigns also made possible the intercomparison of measurements of abundant coma species across spectra ranging from the ultraviolet, infrared, and sub-millimeter.  \citet{Feaga:2014} have presented a compendium of measurements tracing the production of \Htwoo\ and CO over time.

A second target-of-opportunity comet, also with a very rich volatile composition, C/2014 Q2 (Lovejoy) \citep{Biver:2016}, was observed in February 2015.  A separate program was devoted to the Sun-grazing comet C/2012 S1 (ISON), which was observed a few weeks before it disintegrated during its close approach to the Sun.  In this paper, we will focus on the two brightest comets, Garradd and Lovejoy, and discuss the observed COS spectra and their analyses.  In both comets, nineteen bands of the CO Fourth Positive system are clearly identified, together with strong partially resolved multiplets of \ion{S}{1}, centered at 1429 \AA\ and 1479 \AA, and multiplets of \ion{C}{1} at 1561 \AA\ and 1657 \AA.  At the 1~\AA\ resolution of COS, the \ion{S}{1} $^3$P -- $^3$D$^o$ multiplet at 1479 \AA\ is resolved from the overlapping intercombination $^3$P -- $^5$D$^o$ multiplet, both of which can be modeled by resonance fluorescence of solar radiation.  
Residual features in the spectrum of comet Garradd are identified as due to terrestrial \ion{He}{1} emission in third order, while in the spectrum of comet Lovejoy they are due to solar Ly$\alpha$ and Ly$\beta$ pumped fluorescence of \Htwo.
The CO column density in the COS aperture is derived from the CO Fourth Positive spectrum using the fluorescence model of \citet{Lupu:2007}, and translated into a mean CO production rate assuming uniform radial outflow.  For context, the comet's water production rate in each comet is derived from nearly concurrent STIS spectra of the OH \asig\ -- \xpi\ (0,0) band at 3085 \AA, which enables calculation of the relative CO/\Htwoo\ abundance. 

\section{OBSERVATIONS and DISCUSSION}

A log of the COS observations is given in Table~\ref{obs}.  For COS, the G160M grating and $2.\!''5$ diameter aperture were used.  In addition, for OH context, observations were made with STIS using the G230L grating and the $0.\!''2 \times 25''$ slit, usually within two hours of the COS observation.

\begin{deluxetable*}{lcccccc}[h]
\tabletypesize{\small}
\tablewidth{0pt}
\tablecaption{\hst/COS Observations of Comets. \label{obs}}
\tablehead{
\colhead{Comet} & \colhead{Date} & \colhead{Start Time (UT)} & \colhead{$r_h$ (au)} & \colhead{$\dot{r}_h$ (\kms)} & \colhead{$\Delta$ (au)} &  \colhead{COS Exposure Time (s)}}
\startdata
103P/Hartley 2 & 2010 Sep 25 & 01:51:11 & 1.15 & --9.3 & 0.22  & 2843 \\
103P/Hartley 2 & 2010 Nov 04 & 05:40:51 & 1.06 & +2.1 & 0.15  & 2410 \\
103P/Hartley 2 & 2010 Nov 28 & 16:01:41 & 1.14 & +9.0 & 0.27 & 2371 \\
C/2009 P1 (Garradd) & 2012 Jan 19 & 03:38:55 & 1.59 & +5.5 & 1.72 & 1000 \\
C/2012 S1 (ISON) & 2013 Oct 08 & 20:18:29 & 1.50 & --34.3 & 1.92 &2728 \\
C/2012 S1 (ISON) & 2013 Oct 21 & 20:56:06 & 1.23 & --37.7 & 1.53 &2563 \\
C/2012 S1 (ISON) & 2013 Nov 01 & 15:39:36 & 0.99 & --42.0 & 1.22 & 2364 \\
C/2014 Q2 (Lovejoy) & 2015 Feb 02 & 21:27:46 & 1.29 & +1.2 & 0.81 & 2999 \\
\enddata
\end{deluxetable*}

\subsection{Comet C/2009 P1 (Garradd)}

\subsubsection{Spectra}

\begin{figure*}[ht]
\begin{center}
\includegraphics*[width=0.8\textwidth,angle=0.]{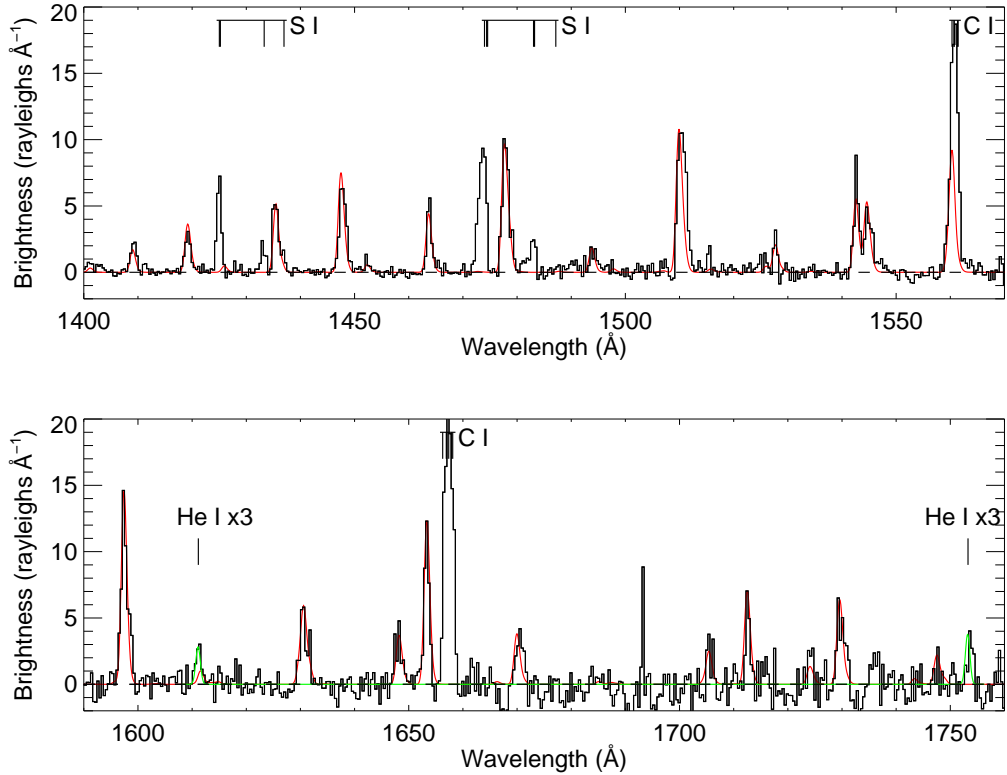}
\caption {COS G160M spectrum of comet C/2009 P1 (Garradd).  Top panel: FUV B channel; bottom panel: FUV A channel. The CO Fourth Positive band synthetic spectrum is overplotted in red.  The multiplets of atomic C and S are indicated.  Also identified are two \ion{He}{1} lines (green) due to terrestrial emission, seen in third order.  \label{cos_spec}}
\end{center}
\end{figure*}

The spectrum of comet C/2009 P1 (Garradd) is shown in Fig.~\ref{cos_spec}, together with the synthetic CO Fourth Positive band model described below, convolved to the 1~\AA\ spectral resolution of COS.  
Although not a point source, the comet's CO emission should be strongly peaked at the nucleus, and we use gaussian convolution to model the shape of the spectral lines.  In addition, the COS line spread function for a point source also deviates from a gaussian\footnote{\tt Fox, A. J., et al. 2015, ÒCOS Data HandbookÓ, Version 3.0, (Baltimore: STScI)}.  These effects can be seen in the detailed spectra around the atomic multiplets as shown in Section~\ref{atoms}.  There is also vignetting around the perimeter of the circular aperture that can distort the fluxes extracted by the calibration pipeline \citep{Cunningham:2015}.  Fig.~\ref{cos_spec} also identifies third order emissions of \ion{He}{1} at 537 and 584 \AA\ that are terrestrial, not cometary, in origin.

\begin{figure}[ht]
\begin{center}
\includegraphics*[width=0.45\textwidth,angle=0.]{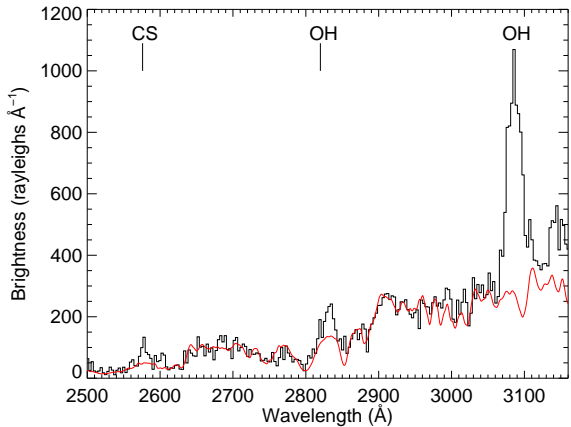}
\caption {Long wavelength region of the STIS G230L spectrum of comet C/2009 P1 (Garradd) obtained 2~hours following the COS observation shown in Fig.~\ref{cos_spec}.  The spectrum is extracted from the $0.\!''5$ central region of the $0.\!''2$ wide slit.  The red line is a solar spectrum convolved to the resolution of STIS. \label{stis_spec}}
\end{center}
\end{figure}

The long wavelength region of the STIS G230L spectrum taken 2~hours following the COS spectrum shown in Fig.~\ref{cos_spec} is shown in Fig.~\ref{stis_spec}.  Besides the known bands of OH and CS, the spectrum is that of solar reflected radiation from cometary dust.

\subsubsection{Modeling \label{model}}

We fit the CO Fourth Positive bands using an optically thick model, following the approach of \citet{Lupu:2007} for treating saturation and self-absorption of the individual lines in a ro-vibrational band, which become significant at CO column densities of a few $\times 10^{14}$~cm$^{-2}$.  The average brightness in the COS $2.\!''5$ diameter aperture is modeled by dividing the circular aperture into five uniformly spaced annuli with relative column densities for spherical radial outflow, calculating a synthetic spectrum for each annulus, and then adding them together in proportion to the area of each annulus.  From the best fit to the observed spectrum (see Fig.~\ref{cos_spec}) the mean column density in the full aperture is obtained and the CO production rate is evaluated assuming radial outflow from a symmetric point source.  Adjustable parameters include rotational temperature ($T_{rot}$), outflow velocity, and solar flux.  We use $T_{rot} = 50$~K \citep{Paganini:2012}, an outflow velocity of $0.85 r_h^{-0.5}$ \kms, consistent with the value derived by \citet{Biver:2012}, and a solar flux appropriate to moderate solar activity.  Because the Fourth Positive bands are pumped mainly by solar continuum, the variation from solar minimum to solar maximum is only $\sim$15\%.  With the uncertainties in the other parameters, our derived value of the mean CO column density in the COS aperture is $1.02 \times 10^{15}$~cm$^{-2}$.  Comparing the model to the observed spectrum, taking into account the nine strongest unblended bands, gives a production rate of $Q$(CO) = $2.16 \pm 0.28 \times 10^{28}$~\mols.

\begin{figure}[ht]
\begin{center}
\includegraphics*[width=0.45\textwidth,angle=0.]{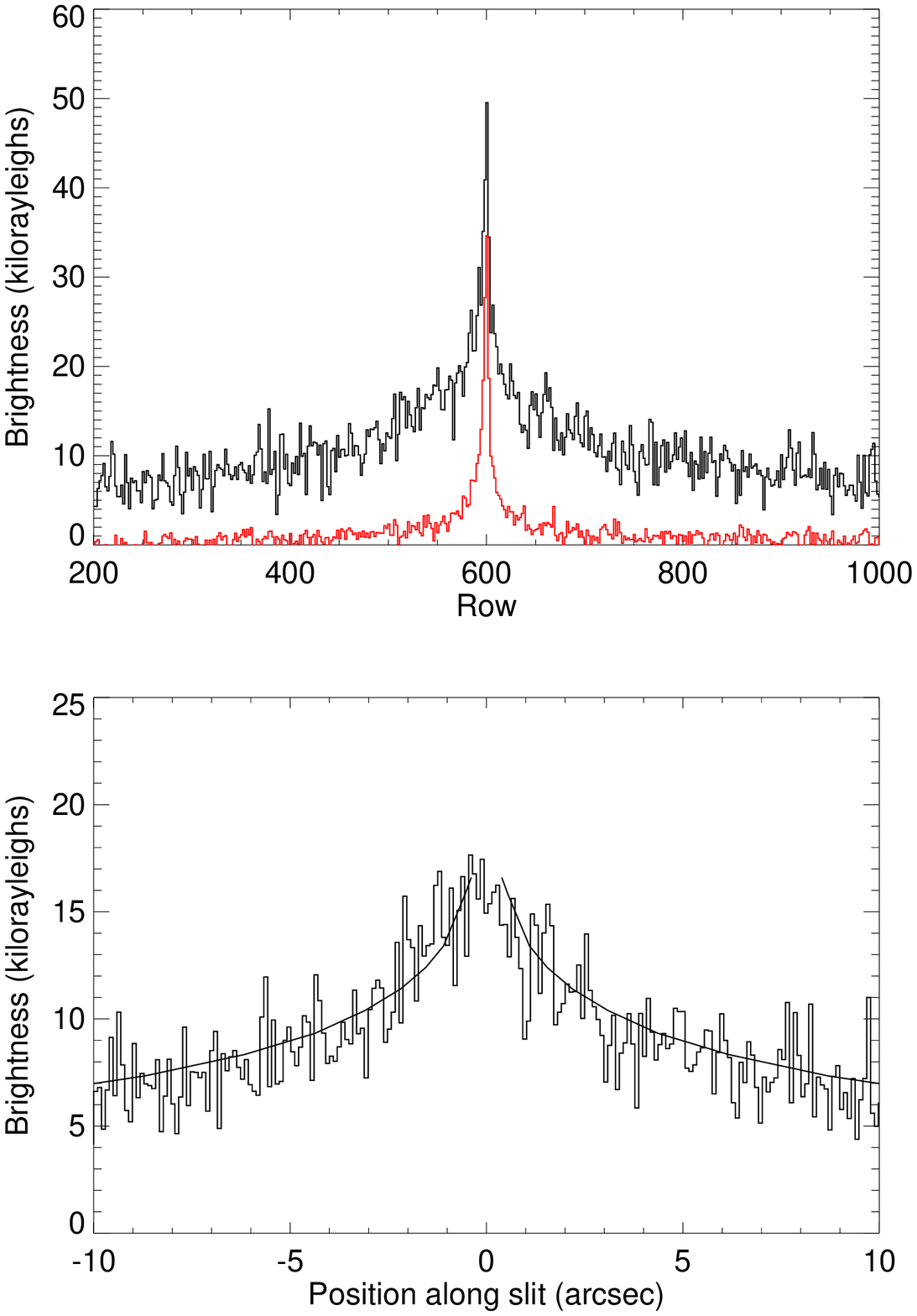}
\caption {Top: Spatial profiles of the OH (0,0) band at 3085~\AA\ and continuum (2860--3050~\AA, red) of comet C/2009 P1 (Garradd), normalized to the solar spectrum in the respective bands.  The lower panel shows a vectorial model fit to the difference profile corresponding to a water production rate of $1.0 \times 10^{29}$ \mols.  One arc second along the slit is 1,250~km at the comet.  \label{stis_prof}}
\end{center}
\end{figure}

From a nearly simultaneous STIS observation of the OH (0,0) band at 3085 \AA, using a vectorial model \citep{Festou:1981} with parameters for moderate solar activity from \citet{Budzien:1994} and a heliocentric velocity dependent OH fluorescence efficiency from \citet{Schleicher:1988} to fit the spatial profile of Fig.~\ref{stis_prof}, yields $Q$(\Htwoo) = $1.0 \times 10^{29}$ \mols.  This is somewhat lower than that derived from {\it SOHO}/SWAN Ly$\alpha$ images from the same day \citep{Combi:2013} or from SWIFT measurements of OH emission \citep{Bodewits:2014}.  Both of these instruments have much larger fields-of-view and are thus sensitive to water from icy particles in the coma, as suggested by \citeauthor{Combi:2013}  We note that over 27~hours the water production rate, as measured by STIS, remained constant to within 10\%. The continuum, however, decreased by a factor of 2 over this interval.  Regrettably, there was only a single COS exposure so that the likely short-term variability of CO seen in other comets \citep{Feldman:2009} could not be determined.
Combining these results for 2012 Jan 19.2 gives a CO/\Htwoo\ production rate ratio of $\sim$22\%. 

From the spatial profile of the CS \api\ -- \xsig\ (0,0) band at 2576 \AA\  seen in Fig.~\ref{stis_spec}, and using the vectorial model with a parent lifetime of 1000~s at 1~au that best fits the profile, we derive a CS parent production rate of 0.1\% relative to \Htwoo.  This value is ÒtypicalÓ for many comets \citep{Meier:1997}.  While the presumed parent, CS$_2$ is less abundant than the other possible direct parent of CS, OCS \citep{Calmonte:2016}, its photodissociation lifetime is considerably shorter than that of OCS \citep{Huebner:2015}, so that CS$_2$ is likely the primary source of the observed CS in the STIS aperture.  We note, however, that \citeauthor{Huebner:2015} recommend a lifetime of $\sim$500~s for CS$_2$, which suggests that the longer-lived species, OCS, may be contributing to the observed spatial profile.
 
\subsubsection{Comparison with other CO observations}

The orbital geometry of comet C/2009 P1 (Garradd) enabled the investigation of its volatile evolution over an extended period around perihelion from both space and ground-based observations.  CO was detected at mm wavelengths in July 2011 at the JCMT \citep{Yang:2012} and in the infrared in September 2011 \citep{Paganini:2012,Villanueva:2012}.  These pre-perihelion observations near 2~au gave CO production rates of the order of $1 \times 10^{28}$~\mols\ or about 10\% that of water and led to the identification of this comet in the class of CO-rich comets.  
\citet{Feaga:2014}, summarizing further JCMT measurements, give $Q$(CO) = $1.3^{+1.0}_{-0.4} \times 10^{28}$~\mols\ for 2012 January 6-8.
\citeauthor{Feaga:2014}, using the infrared spectrograph on the {\it Deep Impact Flyby} spacecraft on 2012 March 26 and April 2 ($r_h \approx  2$~au), found a value of $Q$(CO) of $2.9 \pm 0.8 \times 10^{28}$~\mols, but more surprisingly that this value was $\sim$60\% that of \Htwoo\ due to the rapidly decreasing water production rate with increasing heliocentric distance.  Our result fits the trend found by \citeauthor{Feaga:2014}, although there are contrary results from millimeter-wave observations \citep{Gicquel:2015} made in the same time period.  For 2012 January 18.63, the day before the \hst/COS observation, \citeauthor{Gicquel:2015} report $Q$(CO) = $1.08 \times 10^{28}$~\mols, a factor of 2 below our \hst\ value.  \citeauthor{Gicquel:2015} also report similar values of $Q$(CO) for January 20.84 and 22.25, and again for dates in February and March, suggesting a constant CO production rate through 2012 March 26 ($r_h =  2.00$~au).  However, the value of $Q$(CO) = $1.01 \pm 0.31 \times 10^{28}$~\mols\ given by \citeauthor{Gicquel:2015} for March 26 is in strong disagreement, within the stated uncertainties, with the result of \citeauthor{Feaga:2014} for this date cited above.  Other post-perihelion CO observations, although not as close in time to the COS observations, include those with IRTF/CSHELL \citep{McKay:2015} that gave $Q$(CO) = $1.64 \pm 0.17 \times 10^{28}$~\mols\ on 2012 January 25, and IRAM millimeter observations \citep{Biver:2012} of $Q$(CO) = $\sim 2.0 \times 10^{28}$~\mols\ on 2012 February 19.

\subsection{Comet C/2012 S1 (ISON)}

Comet C/2012 S1 (ISON) was observed on three distinct occasions in 2013 October and November (Table~\ref{obs}).  Several CO Fourth Positive bands are clearly detected in the spectra from October 21 and November 1, but not from October 8.  For the last of these dates the derived CO column density translates to a CO production rate of $2.6 \pm 0.4 \times 10^{26}$ \mols.  From the STIS OH observation closest in time on that date the derived mean \Htwoo\ production rate is $\sim 2.0 \times 10^{28}$ \mols, 
in good agreement with the time-varying production rate found by \citet{Combi:2014} from $SOHO$/SWAN observations of the Ly$\alpha$ coma.  This gives a ratio $Q$(CO)/$Q$(\Htwoo) $\approx 1.3$\%.  During several days between November 15 and November 22, \citet{DiSanti:2016} found values of $Q$(CO)/$Q$(\Htwoo) close to 1.5\%, suggesting that this abundance ratio remained constant while the overall outgassing rate increased by almost two orders of magnitude over this one month period  \citep{Combi:2014}.

An early attempt to obtain spectroscopic observations of comet ISON with \hst\ on 2013 May 02 at a heliocentric distance of 3.88~au did not result in the detection of any gas emissions.  A full discussion of the COS observations of comet ISON will be presented elsewhere.

\subsection{Comet C/2014 Q2 (Lovejoy)}

\begin{figure*}[ht]
\begin{center}
\includegraphics*[width=0.8\textwidth,angle=0.]{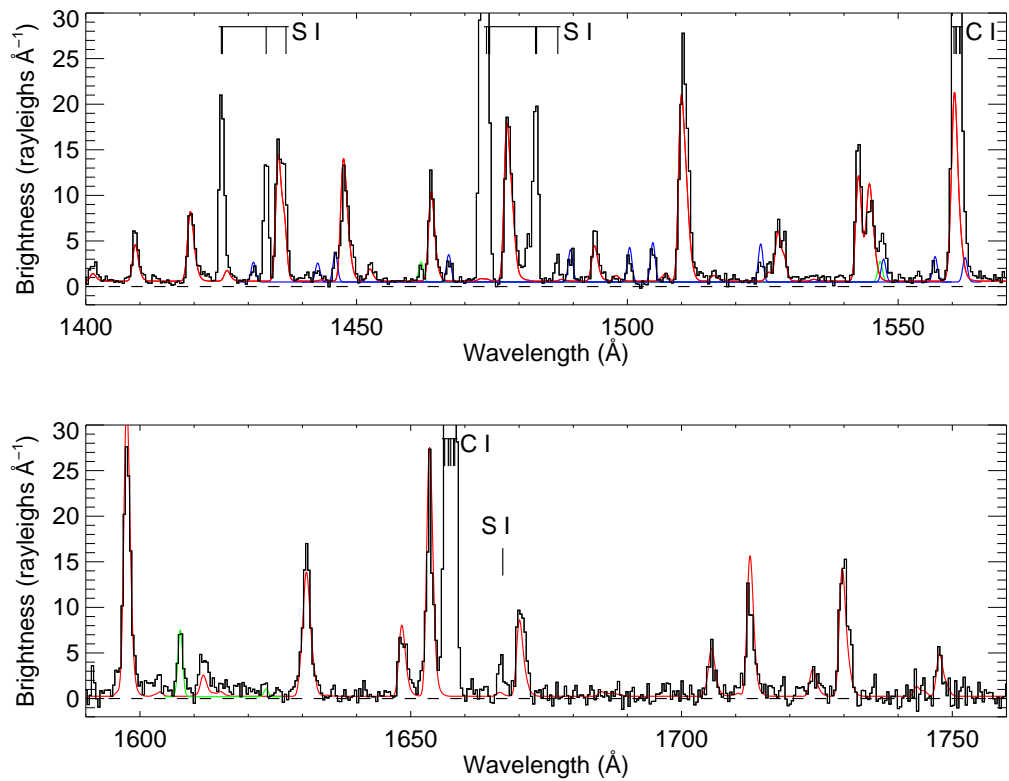}
\caption {The observed spectrum of comet C/2014 Q2 (Lovejoy).  The red line represents a synthetic CO Fourth Positive fluorescence spectrum.  The green lines are solar Ly$\beta$ pumped \Htwo\ fluorescence while the blue lines are a synthetic  spectrum of Ly$\alpha$ pumped \Htwo\ vibrationally excited in the photodissociation of \Htwoco\ \citep{Feldman:2015a}.  \label{lovejoy_spec}}
\end{center}
\end{figure*}

The COS spectrum of comet C/2014 Q2 (Lovejoy) is shown in Fig.~\ref{lovejoy_spec}.  As in the case of comet Garradd, (Fig.~\ref{cos_spec}), 19 bands of the CO Fourth Positive system are clearly detected.  A synthetic model spectrum is similarly overplotted, but with a rotational temperature $T_{rot} = 75$~K, close to the value derived from infrared spectra of \Htwoo\ by \citet{Paganini:2017}.  The derived mean column density is $2.02 \times 10^{15}$~cm$^{-2}$.  The high signal-to-noise ratio of this spectrum allows for the detection of lines of the Lyman band system ($B\,^1\Sigma_u^+ - X\,^1\Sigma_g^+$) of \Htwo, fluorescently pumped from the ground vibrational level by solar Ly$\beta$ \citep{Feldman:2002} (in green) and from excited vibrational levels by solar Ly$\alpha$ (in blue).  The latter result from the photodissociation of \Htwoco\ \citep{Feldman:2015a} and were first identified in spectra of Jupiter following the impact of comet Shoemaker-Levy 9 in 1994 \citep{Wolven:1997}. These are discussed in detail below in Section~\ref{h2}.
 
Using the same approach as in Section~\ref{model}, the derived CO column density translates to a global production rate of $Q$(CO) = $2.23 \pm 0.27 \times 10^{28}$~\mols.  From a similar vectorial analysis of the STIS observation of the OH (0,0) band, we find $Q$(\Htwoo) = $4.5 \times 10^{29}$ \mols, again with an uncertainty of $\sim$20\%.  The derived water production rate can be compared with several in the literature made during the same time frame.  From high resolution infrared spectroscopy, \citet{Faggi:2016} found $Q$(\Htwoo) = $5.0 \times 10^{29}$ \mols\ on February 2, while \citet{Paganini:2017} found $Q$(\Htwoo) = $5.9 \times 10^{29}$ \mols\ on February 4.  Both of these required significant aperture corrections.  \citet{Combi:2018}, again from $SOHO$/SWAN observations of the Ly$\alpha$ coma on February 2.0, find the same value as \citeauthor{Paganini:2017}  However, \citet{Biver:2016}, from observations with the {\it Odin} submillimeter satellite between January 29 and February 3, showed that $Q$(\Htwoo) was variable with an approximate 1~day period and an amplitude variation of $\pm$20\%, and the \hst /STIS result is consistent with this temporal behavior.  Thus, for comet Lovejoy we find $Q$(CO)/$Q$(\Htwoo) $= 5.0$\%.

This result differs from those from millimeter-wave CO observations made a few weeks prior to perihelion.  \citet{deVal-Borro:2018} found, from observations on 2015 January 17-18, a CO production rate of $1.2 \pm 0.3 \times 10^{28}$~\mols\ and a ratio $Q$(CO)/$Q$(\Htwoo) $= 2.0$\%. \citet{Biver:2015} only quoted a ratio of 1.8\% from an average of observations made 2015 January 13-16 and January 23-25, which is in agreement with \citeauthor{deVal-Borro:2018}
The difference may reflect an effect due to the large difference in field-of-view between COS and the ground-based millimeter telescopes, or a true change in outgassing rates with time as in the case of comet Garradd.

As in Section~\ref{model}, from the spatial profile of the CS (0,0) band at 2576 \AA, we derive a CS parent production rate of $4.5 \times 10^{26}$ \mols, which again is 0.1\% relative to \Htwoo.  In this case we can compare with the results of mm observations made between 2015 January 13 and 24. \citet{Biver:2016} give an average production rate for the period of $2.3 \pm 0.02 \times 10^{26}$ \mols, but it is not clear if the discrepancy is due to changes between observing epochs or uncertainties in the parameters used to model the observations.  

\section{ATOMIC EMISSIONS \label{atoms}}

\subsection{\ion{S}{1} $\lambda\lambda$1429 and 1479}

\begin{figure}[ht]
\begin{center}
\includegraphics*[width=0.47\textwidth,angle=0.]{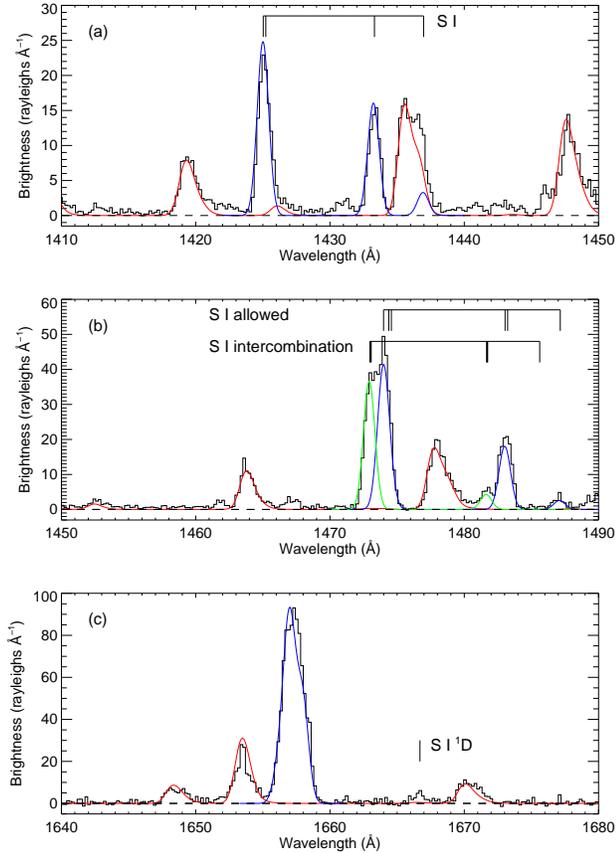}
\caption {Details of the \ion{S}{1} and \ion{C}{1} emission multiplets from the spectrum of comet Lovejoy (Fig.~\ref{lovejoy_spec}).  The synthetic CO spectrum is again shown in red, while model fits to the allowed transitions are shown in blue.  In (b), modeled emission of the intercombination \ion{S}{1}] multiplet is shown in green. The position of the  \ion{S}{1} $^1$D -- $^1$D$^o$ line at 1667~\AA, expected from electron dissociative excitation of SO$_2$, CS$_2$, or H$_2$S, is also shown in (c).  \label{cos_atoms}}
\end{center}
\end{figure}

The \ion{S}{1} multiplets centered at 1429 \AA\ and 1479 \AA\ were first discussed by \citet{Roettger:1989} in the context of \iue\ observations of comet C/1986 P1 (Wilson).  In contrast to the resonance triplet \ion{S}{1} $\lambda\lambda$1807, 1820 and 1826 \citep{Meier:1997}, \citeauthor{Roettger:1989} noted that these multiplets were detected only in comets with small heliocentric velocity where resonance scattering of the narrow solar \ion{S}{1} lines was possible.  Long-slit spectroscopy also showed that all of the \ion{S}{1} emissions were very extended \citep{McPhate:1999}, confirming that they were produced by resonance scattering of S atoms following photodissociation of short-lived sulfur-bearing molecules and not by direct excitation of these same molecules.
At the 1 \AA\ resolution of COS the lines around 1479~\AA\ are partially resolved into the allowed $^3$P -- $^3$D$^o$ multiplet and the semi-forbidden intercombination $^3$P -- $^5$D$^o$ multiplet (Fig.~\ref{cos_atoms}).  These two multiplets were discussed by \citet{Feaga:2002} in the context of \hst /STIS spectral images of Io, where they were also resolved, and attributed to resonance scattering, becoming optically thick at the column densities through the atmosphere of Io.

The relative line intensities can be modeled with solar resonance fluorescence assuming thermal populations of the ground state $J$ levels and treating self-absorption and saturation the same way that \citet{Lupu:2007} did for the CO Fourth Positive bands.  Since the solar \ion{S}{1} lines are narrow, we use high resolution solar spectra from the Ultraviolet Spectrometer Polarimeter Experiment on the {\it Solar Maximum Mission} \citep{Tandberg-Hanssen:1981}, scaled to match the whole disk solar flux for the observation date obtained from {\it TIMED}/SEE measurements \citep{Woods:2005}. The solar spectrum is shifted according to the comet motion relative to the Sun \citep[Swings effect,][]{Swings:1941}.  Oscillator strengths and transition probabilities were taken from the recent compilation of \citet{Deb:2008}. Examples of velocity dependent fluorescence efficiencies (or ``g-factors") are given in Figure~4 of \citet{McCandliss:2016}. 

For comet Lovejoy, both allowed and intercombination multiplets can be fit with a sulfur column density of $1.2 \times 10^{14}$~cm$^{-2}$ as illustrated in Fig.~\ref{cos_atoms}.  The relative line intensities within the multiplets constrain the Boltzmann population of ground state $J$ levels to $T \sim 300$~K. Similar fits are found for the other comets and the derived column densities are tabulated in Table~\ref{qs}.  From the CS analyses discussed above, it is clear that CS is insufficient as the major source of the observed sulfur atoms, but that they must come from the second dissociation of the products of the abundant coma species, H$_2$S and SO$_2$ \citep{Biver:2015}. 
 
\begin{figure}[ht]
\begin{center}
\includegraphics*[width=0.45\textwidth,angle=0.]{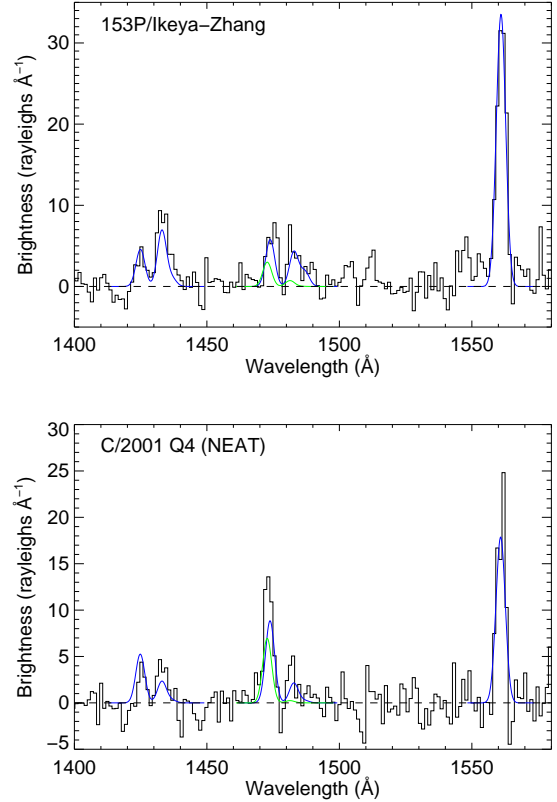}
\caption {Residuals from \hst/STIS spectra of comets 153P/Ikeya-Zhang and C/2001 Q4 (NEAT) after subtraction of CO Fourth Positive band models (from Fig.~7 of \citet{Lupu:2007}), together with models of the \ion{S}{1} and \ion{C}{1} multiplets (blue).  The intercombination multiplet of \ion{S}{1} is shown in green.  The spectrum is integrated over the central $4''$ of the $0.\!''2 \times 25''$ slit and the spectral resolution is 4~\AA.  The \ion{S}{1} multiplets show the effect of the different heliocentric velocity of the two comets (Table~\ref{qs}).  \label{iz_q4}}
\end{center}
\end{figure}

Electron dissociative excitation of sulfur-bearing molecules such as SO$_2$, CS$_2$, or H$_2$S, can also produce the observed \ion{S}{1} emissions \citep{VattiPalle:2004,Brotton:2009,Brotton:2011b}.  Electron excitation of an analogous intercombination \ion{O}{1} multiplet in the innermost coma of comet 67P/Churyumov-Gerasimenko observed by {\it Rosetta} was reported by \citet{Feldman:2015b} and SO$_2$ and H$_2$S were among the most abundant sulfur-bearing molecules recorded by the mass spectrometer experiment on this mission \citep{Calmonte:2016}.  However, if this mechanism was significant at the scale of the COS observations ($\sim$500~km), from the laboratory data we would expect to see several tens of rayleighs of \ion{S}{1} ($^1$D -- $^1$D$^o$) emission at 1666.9 \AA, which is only weakly present in the spectrum of comet Lovejoy (Fig.~\ref{cos_atoms}c) and which we attribute to resonance scattering of sunlight from the short-lived lower state of the transition.  

\subsection{\ion{C}{1} $\lambda\lambda$1561 and 1657}

Unlike the \ion{S}{1} multiplets, the two \ion{C}{1} multiplets do not separate into three groups resolvable at the spectral resolution of COS.  Nevertheless, they can be modeled in the same way permitting a determination of the atomic carbon column density, as given in Table~\ref{qs}.  The fit to the envelope of the \ion{C}{1} $\lambda$1657 multiplet in comet Lovejoy is shown in Fig.~\ref{cos_atoms}c.

\subsection{Comparison with \hst\ observations of other comets}

Two comets observed with STIS in the same spectral region are listed in Table~\ref{qs}.  Figure~7 of \citet{Lupu:2007} shows the  \ion{S}{1} and \ion{C}{1} emissions in the residuals from STIS spectra of comets 153P/Ikeya-Zhang and C/2001 Q4 (NEAT) after subtraction of CO Fourth Positive band models.  At the $\sim$4~\AA\ resolution of the STIS G140L grating the \ion{S}{1} multiplets are partially resolved and the Swings effect is clearly seen.  Fig.~\ref{iz_q4} shows the fit of our models to the residual spectra.  The derived atomic column densities are also given in Table~\ref{qs}.

\begin{deluxetable*}{lccccccccc}
\tabletypesize{\small}
\tablewidth{0pt}
\tablecaption{Production Rates and Column Densities. \label{qs}}
\tablehead{
\colhead{Comet} & \colhead{Date} & \colhead{$r_h$} & \colhead{$\dot{r}_h$} & \colhead{$\Delta$} &  \colhead{Q(CO)} & \colhead{Q(\Htwoo)} & \colhead{Q(CO)/Q(\Htwoo)} & \colhead{N(C)}  & \colhead{N(S)} \\
\colhead{} & \colhead{} & \colhead{(au)} & \colhead{(\kms)} & \colhead{(au)} &  \multicolumn{2}{c}{($\times 10^{28}$~\mols)} & \colhead{} & \multicolumn{2}{c}{($\times 10^{13}$ cm$^{-2}$)} }
\startdata
103P/Hartley 2 & 2010 Nov 04 & 1.06 & 2.1 & 0.15 & $2.6 \pm 1.3 \times 10^{-3}$ & 0.85 & 0.003 & 0.03 & 1.2  \\
C/2009 P1 (Garradd) & 2012 Jan 19 & 1.59 & 5.5 & 1.72 & $2.16 \pm 0.28$ & 10. & 0.22 & 0.30 & 1.5 \\
C/2012 S1 (ISON) & 2013 Nov 01 & 0.99 & --42.0 & 1.22 & $2.6 \pm 0.4 \times 10^{-2}$ & 2.0 & 0.013 & 0.05 & 0.6  \\
C/2014 Q2 (Lovejoy) & 2015 Feb 02/03 & 1.29 & 1.2 & 0.81 & $2.23 \pm 0.27$ & 45. & 0.050  & 1.0 & 12.0  \\
\tableline\tableline
153P/Ikeya-Zhang & 2002 Apr 20 & 0.90 & 29.1 & 0.43 &  $1.54 \pm 0.09$ & 21.5 & 0.072 & 6.0 & 20.0 \\
C/2001 Q4 (NEAT) & 2004 Apr 25 & 1.02 & -10.3 & 0.47 &  $1.76 \pm 0.16$ & 20. & 0.088 &1.0 & 4.0 \\
\enddata
s\end{deluxetable*}

\section{MOLECULAR HYDROGEN \label{h2}}

The residual spectrum of \citeauthor{Lupu:2007} of comet 153P/Ikeya-Zhang also showed the presence of the (6, 13) P1 line of the \Htwo\ Lyman band system ($B\,^1\Sigma_u^+ - X\,^1\Sigma_g^+$) at 1607.5~\AA.  This is the strongest line in the (6, $v''$) P1 progression that is pumped from the ground vibrational level of \Htwo\ by the coincidence of solar Ly$\beta$ with the (6, 0) P1 line, the (6, 1), (6, 2), and (6, 3) lines having been detected at wavelengths shortward of Ly$\alpha$ in comet C/2001 A2 (LINEAR) by the {\it Far Ultraviolet Spectroscopic Explorer} (\fuse) \citep{Feldman:2002}.  Further analysis of the \fuse\ spectra of four comets \citep{Feldman:2009,Feldman:2015a} revealed fluorescence from excited vibrational levels of \Htwo\ pumped by solar \ion{O}{6} $\lambda$1031.9 and \ion{H}{1} Ly$\alpha$, which together with a non-thermal energetic population of rotational levels of CO, also detected by \fuse, was identified as produced by photodissociation of \Htwoco\ on the basis of laboratory spectra.  Longer wavelength lines of these progressions are detected in the spectrum of comet Lovejoy (Fig.~\ref{lovejoy_spec}).  A synthetic spectrum of the Ly$\alpha$ pumped lines, using the branching ratios derived by \citet{Wolven:1997}, is overplotted, and the strongest unblended lines are listed in Table~\ref{h2lines}.  The table also lists the fluorescence efficiencies (g-factors) calculated using solar 
Ly$\alpha$ and Ly$\beta$ fluxes and line shapes from \citet{Lemaire:2015} and oscillator strengths and branching ratios derived from the transition probabilities of \citet{Abgrall:1993a}, together with the column densities derived from the observed line brightnesses.

\begin{deluxetable*}{ccccc}
\tabletypesize{\small}
\tablewidth{0pt}
\tablecolumns{5}
\tablecaption{Strongest unblended \Htwo\ lines in comet C/2014 Q2 (Lovejoy). \label{h2lines}}
\tablehead{
\colhead{Wavelength} & \colhead{Transition} & \colhead{Brightness} & \colhead{g-factor at 1~au} & \colhead{\Htwo($v'',J''$) column} \\
\colhead{(\AA)} & \colhead{} & \colhead{(rayleighs)} & \colhead{(photons~s$^{-1}$~molecule$^{-1}$)} & \colhead{density (cm$^{-2}$)}  }
\startdata
\multicolumn{5}{l}{Ly$\alpha$ pumping of \Htwo($v''=2, J''=5$)}  \\
1431.01	 & (1, 6) R3  & 	1.67 & $9.60 \times 10^{-6}$ &  $2.89 \times 10^{11}$  \\
1489.57	 & (1, 7) R3  & 	3.46 & $1.69 \times 10^{-5}$ &  $3.39 \times 10^{11}$  \\
1504.76	 & (1, 7) P5  & 	3.95 & $2.10 \times 10^{-5}$ &  $3.13 \times 10^{11}$  \\
\hline
\multicolumn{5}{l}{Ly$\alpha$ pumping of \Htwo($v''=2, J''=6$)}  \\
1467.08	 & (1, 6) P8  & 2.45 & $2.10 \times 10^{-5}$ &  $1.94 \times 10^{11}$	 \\
1500.45	 & (1, 7) R6  & 2.51 & $2.77 \times 10^{-5}$ &  $1.51 \times 10^{11}$	 \\
1524.65	 & (1, 7) P8  & 2.52 & $3.15 \times 10^{-5}$ &  $1.33 \times 10^{11}$	 \\
1556.87	 & (1, 8) R6  & 2.37 & $2.19 \times 10^{-5}$ &  $1.80 \times 10^{11}$ \\
\hline
\multicolumn{5}{l}{Ly$\beta$ pumping of \Htwo($v''=0, J''=1$)}  \\
1607.50	 & (6, 13) P1  &  9.62 & $6.27 \times 10^{-7}$ &  $2.55 \times 10^{13}$ \\
\hline
\enddata
\end{deluxetable*}

The principal photodissociation channels of \Htwoo\ are
\[ \Htwoo\ + h\nu  \rightarrow {\rm H + OH}  \]
and
\[ \Htwoo\ + h\nu  \rightarrow \Htwo\ + {\rm O(^1D)}  \]
\citet{Huebner:2015} give the ratio of photodissociation rates for these two processes as 0.058--0.084, for quiet Sun and active Sun conditions, respectively.  We take the \Htwo\ ($v=0,J=1$) column density from the last line of Table~\ref{h2lines} and derive a total \Htwo\ column density by recognizing that the population of the $J = 1$ level ranges between 0.60 and 0.70 for rotational temperatures between 100 and 300~K.  Thus, $N(\Htwo) = 3.9 \times 10^{13}$~cm$^{-2}$.  We can similarly derive a column density for OH from the STIS spectrum, even though the effective aperture (we take $0.\!''2 \times 2.\!''5$) is not the same as the circular COS aperture.  With this caveat, we find $N({\rm OH}) = 5.1 \times 10^{14}$~cm$^{-2}$, and $N(\Htwo)/N({\rm OH}) = 0.076$, in excellent accord the the relative photodissociation rates.  

Similarly, we can take the photodissociation rate for the process
\[ \Htwoco\ + h\nu  \rightarrow \Htwo\ + {\rm CO}  \]
given by \citet{Huebner:2015} to estimate the amount of \Htwo\ produced this way relative to that from \Htwoo.  We take $Q(\Htwoco)/Q(\Htwoo) = 0.0036$ from \citet{Biver:2016} and find $N(\Htwo)$ from \Htwoco\ varies from 1.1 to $2.7 \times 10^{13}$~cm$^{-2}$, again for active or quiet Sun conditions, respectively.  From Table~\ref{h2lines} we find the average column density of \Htwo\ ($v=2,J=5$) is $3.1 \times 10^{11}$~cm$^{-2}$, and that the population of $v=2,J=5$ ranges from 0.012 to 0.028.  For two comets observed by \fuse, with similar \Htwoco\ production rate relative to \Htwoo, \citet{Feldman:2015a} found the population of this level to be 0.014 and 0.018.  Thus, we can confirm \Htwoco\ as the source of the Ly$\alpha$ pumped lines in the spectrum of comet Lovejoy, even though we cannot unambiguously detect the non-thermal wings, separated by 1.5~\AA\ from the band origin, on the CO Fourth Positive bands.  From the \fuse\ data of \citet{Feldman:2009} and \citet{Feldman:2015a}, and using the mean of the \Htwo\ populations given in Table~\ref{h2lines}, we expect the non-thermal CO population, peaking at $J=40$, to be $3.1 \pm 0.7$\% that of the total CO population, which is consistent with the spectrum shown in Fig.~\ref{lovejoy_spec}.

\section{SUMMARY}

We report on observations of four comets made by the Cosmic Origins Spectrograph on the {\it Hubble Space Telescope} between 2010 and 2015.  In each comet the CO column density is derived from models of the observed Fourth Positive bands and used to derive CO production rates. Nearly concurrent STIS observations of the OH (0,0) band at 3085 \AA, together with a vectorial model give the \Htwoo\ production rate and the abundance of CO relative to \Htwoo\ which ranged from $\sim$0.3\% to $\sim$22\%. The \ion{S}{1} intercombination multiplet, $^3$P -- $^5$D$^o$], at 1479 \AA, is resolved from the allowed resonance multiplet, $^3$P -- $^3$D$^o$ in all four comets and can be satisfactorily modeled by resonance fluorescence of solar radiation by the sulfur atoms in the coma.  In comet Lovejoy we have detected emission from several lines of the \Htwo\ Lyman band system, excited by solar Ly$\alpha$ and Ly$\beta$ pumped fluorescence, and have demonstrated that the brightness of the Ly$\alpha$ pumped lines is consistent with photodissociation of \Htwoco\ into vibrationally excited \Htwo\ levels.

\newpage
\acknowledgments

We thank Alison Vick, Dave Sahnow, Tony Roman, and Tracy Ellis at STScI for their planning efforts and Jon Giorgini at JPL for ephemeris support.    This work is based on observations made with the NASA/ESA Hubble Space Telescope, which is operated by the Association of Universities for Research in Astronomy, Inc., under NASA contract NAS 5-26555.  The work at Johns Hopkins University was supported by NASA through grants HST-GO-12607.003-A, HST-GO-13492.004-A, and HST-GO-13874.003-A, from the Space Telescope Science Institute.

\facility{HST}



\end{document}